\documentclass[11pt]{article}
\pdfoutput=1
 \usepackage{mciteplus}
 \usepackage{tikz}
 \usepackage{color}
 \usepackage{xcolor}
 \usepackage{comment}
 \definecolor{darkblue}{rgb}{0.1,0.1,.7}
 \usepackage[colorlinks, linkcolor=darkblue, citecolor=darkblue, urlcolor=darkblue, linktocpage,hyperfootnotes=false]{hyperref} 
\usepackage{epsfig}
\usepackage{graphicx}
\usepackage{cite}
\usepackage{amsfonts}
\usepackage{amssymb}
\usepackage{bm}
\usepackage{latexsym}
\usepackage{mathtools}
\usepackage{amsmath}
\def\bq{\begin{quote}}
\def\eq{\end{quote}}
 at 10truept

\newcommand{\cala}{{\cal A}}

\newcommand{\calp}{{\cal P}}

\newcommand{\beq}{\begin{equation}}
\newcommand{\eeq}{\end{equation}}
\newcommand{\beqa}{\begin{eqnarray}}
\newcommand{\eeqa}{\end{eqnarray}}
\newcommand{\bea}{\begin{eqnarray}}
\newcommand{\eea}{\end{eqnarray}}

\def\roughly#1{\raise.3ex\hbox{$#1$\kern-.75em\lower1ex\hbox{$\sim$}}}

\begin{document}

\thispagestyle{empty}
\pagenumbering{Alph}
\begin{titlepage}
  \bigskip

  \bigskip\bigskip

  \bigskip

\begin{center}
{\Large \bf {Gravitational dressing: from the crossed product to more general algebraic and mathematical structure
}}
    \bigskip
\bigskip
\end{center}

  \begin{center}

 \rm {Steven B. Giddings\footnote{\texttt{giddings@ucsb.edu}} }
  \bigskip \rm
\bigskip

{Department of Physics, University of California, Santa Barbara, CA 93106, USA}  \\
\rm

  \bigskip \rm
\bigskip
 
\rm

\bigskip
\bigskip

  \end{center}

\vspace{3cm}
  \begin{abstract}
  
  The crossed product, and consequent transition from von Neumann algebras of type III to II, is recovered from a truncation of more general gravitational dressing constructions, about certain spacetimes.  This is done by extending ``standard dressing" constructions previously used to give a perturbative definition of ``gravitational splittings," defining approximate localization of information. This result appears to illustrate that this algebraic transition is a small piece of a more general algebraic, or other mathematical, structure associated with quantum gravity.  The leading-order structure involves noncommutativity from separated regions, and at the nonperturbative level connects with a possible explanation of holographic behavior for gravity.

 \medskip
  \noindent
  \end{abstract}

  \end{titlepage}
\pagenumbering{arabic}


In the search for a quantum theory of gravity, two apparently key questions are to describe observations and observables; these questions are of course related.
Study of properties of quantum observables, which are certain operators on the quantum space of states, is anticipated to in particular yield information about the mathematical structure of quantum gravity, much as study of such observables in local quantum field theory (LQFT) plays a central role in characterizing such theories in the algebraic approach\cite{Haag}.  In gravity, due to the gauge invariance, observables are not fundamentally local\cite{Torr}, and apparently need to be formulated relationally, or asymptotically.  Two kinds of such observables are the field-dressed observables, which are defined relationally with respect to a configuration of matter fields (such as an actual observer), and the gravitationally-dressed observables, which involve a minimal amount of extra gravitational structure to make them gauge invariant.

The latter gravitationally-dressed observables are in particular expected, due to their minimal nature, to be likely to play a role in the basic mathematical structure of the theory.  Many questions remain regarding their general  formulation, but there has been some success in defining them and exploring their properties at a leading perturbative order in Newton's constant $G$\cite{SGalg,DoGi1,DoGi2,DoGi3,GiKi,DoGi4,SGsplit,GiWe,SGsub,GiPe2,obs-over}.\footnote{Important precursors to this work are \cite{Heem,KaLigrav}.}  Interestingly, already at this level, one  finds a significant departure from the algebraic structure of LQFT, whose full meaning we seem to just be beginning to understand.  

In related discussion, there has been considerable interest in gravitational modification of algebras in certain contexts\cite{Wittcross,CPW,CLPW} (see {\it e.g.} \cite{KlLe2023,Fewster:2024pur,DEHK-crossed} for followup work) introducing a crossed product structure\cite{Take}, resulting in
 a transition from the type III von Neumann algebras typical of LQFT to type II algebras.  This, for example, allows entropies to be more readily defined.
 But, there has remained a question about the precise connection between these developments and the aforementioned gravitational dressing story.

This short note addresses this and related questions.  In short, the perturbative gravitational dressing furnishes a generalization of Takesaki's crossed product construction, seen via the constructions of perturbative gravitational splittings by what has been called a standard dressing construction\cite{DoGi3,DoGi4,SGsplit}.  Specifically, the crossed product and type II structure will be shown to arise from a {\it truncation} of the full gravitational dressing, in certain contexts.  This seems to make clear that the crossed product construction and type II algebras are just a small part of an even more interesting and nontrivial broader algebraic structure of quantum gravity.  This note demonstrates this by giving a more general version of standard dressing constructions, based on a formulation of gravitational dressings perturbatively about general backgrounds\cite{GiPe2}.  In terms of this construction, when for example working perturbatively about an eternal black hole, one finds the crossed product emerge directly from a truncation of the more complete gravitational dressing. The corresponding truncated operators of this crossed product construction are not fully gauge invariant.

This discussion also presents a starting point for other generalizations, and of further investigation of this algebraic structure, which will be briefly described.  Going beyond the perturbative dressing, plausible behavior of fully {\it non-perturbatively} dressed observables raises questions about the need for other mathematical structure, beyond that of algebras, to usefully characterize information and its localization in quantum gravity.

Our starting point is a relatively general description of dressed observables, like that outlined in \cite{obs-over}.  For example, if a LQFT has a local observable $O(x)$, this will not be invariant under a diffeomorphism {\it e.g.} with infinitesimal parameter $\xi$,
\beq\label{scalardiff}
\delta_\xi O(x) = -\xi^\mu \partial_\mu O(x)
\eeq
and so is not a good gauge-invariant observable in the gravitational context where diffeomorphisms are associated with gauge symmetries.  At the classical level, this may be remedied {\it relationally}  by finding suitable scalar functionals $X^A[\phi_I, g_{\mu\nu}] (x)$ of other fields $\phi_I$ and the metric $g_{\mu\nu}$ that  transform as scalars, also as in \eqref{scalardiff}.  If these are chosen so that specifying  $X^A=y^A$ also specifies a spacetime point,
\beq\label{dressinv}
x^\mu=\chi^\mu(y^A)\ ,
\eeq
for  some scalar parameters $y^A$,
the {\it dressed observable}
\beq\label{dressedO}
O(\chi^\mu(y^A)) 
\eeq
is a diffeomorphism-invariant version of the local observable at that point.  We may think of the functionals $X^A$ as providing a reference frame.

An important question is extending such relational constructions to the quantum context.\footnote{In the quantum context, the general relational constructions may also be described in the terminology of quantum reference frames\cite{GHK,DEHK,DHT}.  Related constructions in the language of the ``dressing field method" are described in ref.~\cite{FGR}, and references therein.}  This paper will focus on the case with $X^A[g_{\mu\nu}]$, independent of other fields, which we refer to as {\it gravitationally dressed}, in contrast\cite{GiPe2,obs-over} to $X^A[\phi^I]$, the {\it field dressed} case, or a more general mixed observables.  The gravitationally dressed case arguably represents the minimal structure needed in gravity\cite{obs-over}, since it involves only gravitational degrees of freedom; we expect the corresponding quantum operators to create/annihilate quantum field configurations together with their gravitational fields. 

While finding the full quantum gauge-invariant gravitationally dressed observables, analogous to \eqref{dressedO}, is a challenging problem apparently driving at deep issues in quantum gravity, considerable progress has been made\cite{SGalg,DoGi1,DoGi2,DoGi3,GiKi,DoGi4,SGsplit,GiWe,SGsub,GiPe2,obs-over} in finding the leading perturbative gravitational dressing, expected to be relevant in weak-field ({\it e.g.} long-distance) contexts, and expected to give us important clues about the more complete mathematical structure of quantum gravity.

Specifically, we consider working about a general background metric $g_{\mu\nu}$, with quantum perturbed metric $\tilde g_{\mu\nu} = g_{\mu\nu} +\kappa h_{\mu\nu}$, where $\kappa^2=32\pi G$.  We may quantize in a canonical framework by choosing a spatial slicing and using  ADM variables\cite{ADM}, 
\beq
ds^2= -N^2 dt^2 + q_{ij}(dx^i+N^i dt)(dx^j+N^j dt)\ .
\eeq
It  has been argued\cite{DoGi1,DoGi4,GiPe2} that at the leading perturbative order in $\kappa$ a more general LQFT operator $O$ may be dressed as
\beq\label{genlin}
{\hat O} \simeq e^{i\int d^3x \sqrt q n^\nu V^\mu(x) T_{\mu\nu}} O e^{-i\int d^3x \sqrt q n^\nu V^\mu(x) T_{\mu\nu}} 
\eeq
where $n^\mu=(1,-N^i)/N$ is the normal to the slices, $T_{\mu\nu}$ is the matter stress tensor, and $V^\mu(x)$ is a {\it gravitational dressing} that is a functional of the metric perturbation.  Eq.~\eqref{genlin}  may be thought of as arising from a perturbative expansion of \eqref{dressedO}, if we define 
\beq\label{pertdress}
\chi^\mu(y) = \delta^\mu_A y^A+ V^\mu(y)\ .
\eeq

This leading-order dressing $V^\mu$ can be determined by enforcing gauge invariance of $\hat O$, which may be expressed by the condition that $\hat O$ commutes with the constraints
\beq\label{constraints}
{\cal C}_\mu(x) = \left[ T_{\mu\nu}(x) - \frac{G_{\mu\nu}(x)}{8\pi G}\right]\sqrt q n^\nu\ ,
\eeq
which canonically generate diffeomorphisms.  Solving these conditions was described in \cite{GiPe2}, which constructed general gravitational dressings of the form\footnote{Note minor simplifying changes of notation from that of \cite{GiPe2}.}
\beq\label{genv}
V^\mu(x)=\frac{\kappa}{2} \int d^{3}x' \left[H^\mu_{ij}(x',x) p^{ij}(x')+G^{ij\mu}(x',x) h_{ij}(x')\right]
\eeq
where $p^{ij}$ are the (densitized) conjugates to the metric perturbation with normalization
\beq
[p^{ij}(x),h_{kl}(x')] = -\frac{i}{2}\left(\delta^i_k \delta^j_l+\delta^i_l \delta^j_k\right)\delta^3(x-x')\ .
\eeq
In \eqref{genv},  $H^\mu_{ij}$, $G^{ij\mu}$ are c-number functions, and the resulting exponential in \eqref{genlin} is a Weyl operator for metric perturbations.  The condition that $\hat O$ commutes with the constraints then gives  leading-order equations\cite{GiPe2} which can be simplified to
\bea\label{GFeqns}
\left(L^{jk}- {\cal P}^{jk}\right) H^\mu_{jk}(x,x') -\frac{K_{jk}}{\sqrt q} G^{jk\mu}(x,x') &=&  \frac{\delta^3(x-x')}{\sqrt q\, N}\delta^\mu_0\cr
q_{ik} D_l G^{kl\mu}(x,x') - {\cal Q}_i^{kl}H^\mu_{kl}(x,x') &=&\delta^{3}(x-x') \delta _i^\mu\ .
\eea
Here $L^{ij}$ is a differential operator that acts on symmetric tensors $t_{ij}$ as
\beq
L^{ij} t_{ij} = D^i D^j t_{ij} - R^{ij}_q t_{ij} -D_iD^i(q^{kl}t_{kl})\ ,
\eeq
$D_i$ and $R^{ij}_q$ are the covariant derivative and Ricci curvature arising from the background spatial metric $q$,  $K_{ij}$ is the background extrinsic curvature
\beq
K_{ij}= \frac{1}{2N}\left(-\dot q_{ij} + D_iN_j + D_jN_i\right)\
\eeq
of the slices, 
\beq
 \calp^{ij} = \frac{\kappa^4}{2q}\left[P^{ik}P_k^j - \frac{PP^{ij}}{2} - \frac{q^{ij}}{2}\left(P^{kl}P_{kl} -\frac{P^2}{2}\right)\right]\ ,
 \eeq
 where $P^{ij}$ are the background momenta
 \beq
 P^{ij} = -\frac{\sqrt q}{16\pi G}\left(K^{ij}- q^{ij} K\right)\ ,
 \eeq
and $\cal Q$ is a linear differential operator defined by
\beq
{\cal Q}_i^{jk} h_{jk} = \kappa q_{ij} (\delta_{\kappa h} D_k)P^{jk} + \kappa^2 D_kP^{jk} h_{ij}\ .
\eeq

We therefore see from \eqref{GFeqns} that the functions $H^\mu_{ij}$, $G^{ij\mu}$ are solutions to a general Green function problem.  As with other such problems, different solutions  exist,  here describing different dressings\cite{DoGi1,GiPe2}.  The difference between two such solutions corresponds to a solution of the homogeneous version of the equations \eqref{GFeqns}, which describes a linear propagating perturbation of the gravitational field, {\it i.e.} a gravitational wave, showing that the different dressings are in general physically inequivalent.  The perturbation to the gravitational field created by the dressings, or ``dressing field," can be found by calculating the commutator of the operators $h_{ij}$ and $p^{ij}$ with the operators \eqref{genlin}, analogously to calculations described in \cite{DoGi1,DoGi4}.

\begin{figure}[h]
 	\begin{center}
 		\includegraphics[width=0.35\textwidth]{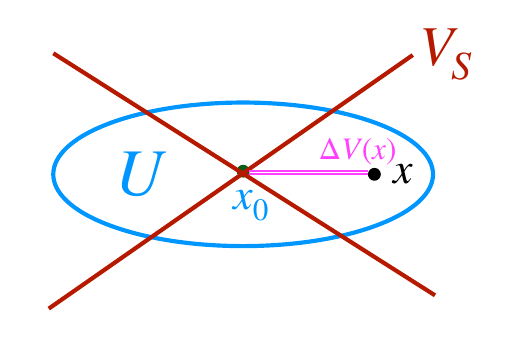} 
 		\caption{Illustration of a standard dressing construction.  The dressing $V_S$ external to the neighborhood $U$ only depends on the neighborhood, {\it e.g.} through choice of a fixed point $x_0$ of the neighborhood.  The full dressing arises by including an additional piece associated with the dressing of a general point.}
 		\label{fig-sdress}
 	\end{center}
 \end{figure} 

Suppose that we consider a subalgebra of LQFT operators $O$ localized in a region $U$.  The resulting dressed observables \eqref{genlin} then do not generally commute for such spacelike separated regions $U$ and $U'$, showing that gravity is responsible for a significant change to the algebraic structure\cite{SGalg,DoGi1,obs-over} of LQFT.  In fact, this raises a question of how information can be localized in quantum gravity, even approximately\cite{DoGi2,DoGi3,DoGi4,SGsplit,QFG,QGQF,SGsub,obs-over}.  At the perturbative level, in flat space, such an approximate answer was provided by the definition of a gravitational splitting\cite{DoGi3,DoGi4,SGsplit}, based on a ``standard dressing" construction. This exploits the existence of different possible dressings. In essence, using this flexibility, the simple gravitational dressings $V^\mu$ considered in that flat context, given a neighborhood $U$, could be divided into pieces,
\beq\label{stdress}
V^\mu(x) = V^\mu_S + \frac{1}{2} x_\nu\left(\partial^\nu V_S^\mu-\partial^\mu V_S^\nu\right) + \Delta V^\mu(x)\ ;
\eeq
this construction is illustrated in Fig.~\ref{fig-sdress}.
Here the standard piece $V^\mu_S$  just depends on the neighborhood, and so is $x$-independent; its derivative is computed by translation of the neighborhood; and for $x\in U$, $\Delta V$ is restricted to the neighborhood.
With this form of the dressing, the expressions \eqref{genlin} simplify to the form
\beq\label{Sdressop}
{\hat O} \simeq \ e^{- iP_\mu V_S^\mu -\frac{i}{2}M_{\mu\nu}\partial^\mu V^\nu_S}\,{\tilde O}\, e^{ iP_\mu V_S^\mu +\frac{i}{2}M_{\mu\nu}\partial^\mu V^\nu_S}
\eeq
where $P_\mu$ and $M_{\mu\nu}$ are the Poincar\'e generators 
\beq
P_\mu=-\int d^3x T_{0\mu}(x)\quad ,\quad M_{\mu\nu}= - \int d^3x  \left[x_\mu T_{0\nu}(x) - x_\nu T_{0\mu}(x)\right]
\eeq
of the underlying LQFT, and $\tilde O$ is a dressed version of $O$ whose dressing $\Delta V$ has support only within $U$.  This shows that measurements of the gravitational perturbation outside $U$ only depend on the state created by $O$ in the region through its Poincar\'e moments\cite{DoGi3,DoGi4,SGsplit,SGsub}, providing an approximate notion of localization of information; they also depend on the standard dressing $V^\mu_S$.  This can be thought of as a linearized quantum gravity version of results of Corvino and Schoen\cite{CoSc} and Carlotto and Schoen\cite{CaSc} showing that outside a region, the classical gravitational field may be taken to be in a standard form, {\it e.g.} the boosted Kerr solution or a cone-localized field.

It seems probable that this gravitational modification of algebraic structure, or perhaps other related mathematical structure, plays an important role in the formulation of a quantum theory of gravity.  One observation in this direction is that the expression \eqref{Sdressop} bears a strong resemblance\cite{obs-over} to the crossed product construction of Takesaki\cite{Take}, which is responsible for a change of algebraic type of von Neumann algebras from III to II, as has been explored in \cite{Wittcross,CPW,CLPW} and many subsequent references; this for example allows definition of entropies.  Specifically, given an algebra $\cal A$ and a group $G$ acting on it as an outer automorphism, the crossed product algebra ${\cal A}\rtimes G$ is generated by
\beq\label{xprodalg}
\{e^{-ix_aH_a}\, {\bf a}\, e^{ix_aH_a}, p_a\}
\eeq
with $\bf a$ an element of $\cal A$, $x_a$ a coordinate on the group $G$, $H_a$ a generator of the $G$-action on $\cal A$, and $p_a$ a conjugate variable to $x_a$.\footnote{These conventions differ with \cite{CLPW,CPW,WittARO} by exchange of  $x$ and $p$.}  

To compare \eqref{xprodalg} with \eqref{Sdressop}, note first that the ADM momenta have commutators with the dressing\cite{DoGi1}
\beq
[P_\mu^{ADM},V_S^\nu]= i \delta_\mu^\nu\ ,
\eeq
so the standard dressings and these momenta enter as $x_a$  and $p_a$ in \eqref{xprodalg};  the angular momenta $M_{\mu\nu}$ behave similarly.   Moreover, in \eqref{Sdressop} the operator $\tilde O$ lies in the algebra of the region $U$, albeit with the inclusion of the dressing $\Delta V$ within that region.  But a significant  difference with \eqref{Sdressop} is that the Poincar\'e generators are not in general automorphisms of the algebra associated with $U$, since for example the operator $e^{ia^\mu P_\mu}$ translates operators by a finite displacement $a^\mu$, and in general outside the region $U$.  
For this reason, \cite{obs-over} argued that gravitational dressing constructions have (re)discovered an interesting generalization of the crossed product construction of Takesaki.

This note will further develop this observation, by further developing and generalizing the standard dressing construction, and by showing that a truncation of the resulting dressed operator algebra reduces to a bona fide crossed product in special cases, connecting with \cite{Wittcross,CPW}.

The first step is to extend standard dressing constructions to the context of more general dressings of \cite{GiPe2} , as briefly reviewed above.  The original standard dressing constructions\cite{DoGi3,DoGi4,SGsplit} directly constructed specific dressings \eqref{stdress} satisfying the relevant conditions for diffeomorphism invariance.  The work of \cite{GiPe2} recasts the problem of finding dressings as that of solving for appropriate Green functions, as described above.  Thus we can extend the standard dressing construction for operators in a localized neighborhood $U$ by finding an appropriate decomposition of the Green functions into pieces corresponding to the different terms of the standard dressing.  For example, in the flat example described above, the standard dressing construction of \cite{DoGi3,DoGi4,SGsplit} can be translated into providing a construction of Green functions that decompose as
\bea
H^0_{ij}(x',x) &=& H^0_{S,ij}(x') + x_\nu\partial^{\nu} H^0_{S,ij}(x') + H^0_{loc,ij}(x',x)\cr
G^{ijk}(x',x) &=& G^{ijk}_S(x') + x_\nu \partial^{\nu}  G_S^{ijk}(x') + G^{ijk}_{loc}(x',x)\ ,
\eea
where $H_S$ and $G_S$ are the pieces associated with the standard dressing.  These are the terms responsible for creating an asymptotic gravitational field which has the necessary ADM charges $P_\mu^{ADM}$ and $M_{\mu\nu}^{ADM}$.   
$H_{loc}$ and $G_{loc}$ are pieces which localize to $U$ for $x,x'\in U$, and in this case $H^k_{ij}$ and $G^{ij0}$ vanish. 

Working about a general background metric $g$, one does not expect such a simple decomposition of the Green functions.  However, for symmetric backgrounds in particular, an interesting question is to what extent do analogous constructions exist.

A next question is that of finding examples where at least part of the group action gives an automorphism of the underlying algebra of observables.  Ref.~\cite{Wittcross,CPW} considered the example of a static black hole spacetime, which has a timelike Killing vector preserving its interior.  This may be taken to lie in either asymptotically flat or AdS space; we focus on the flat case, though extension of the results in \cite{GiKi} regarding dressings and asymptotic charges in AdS is expected to be straightforward and analogous.  

\begin{figure}[h]
 	\begin{center}
 		\includegraphics[width=0.80\textwidth]{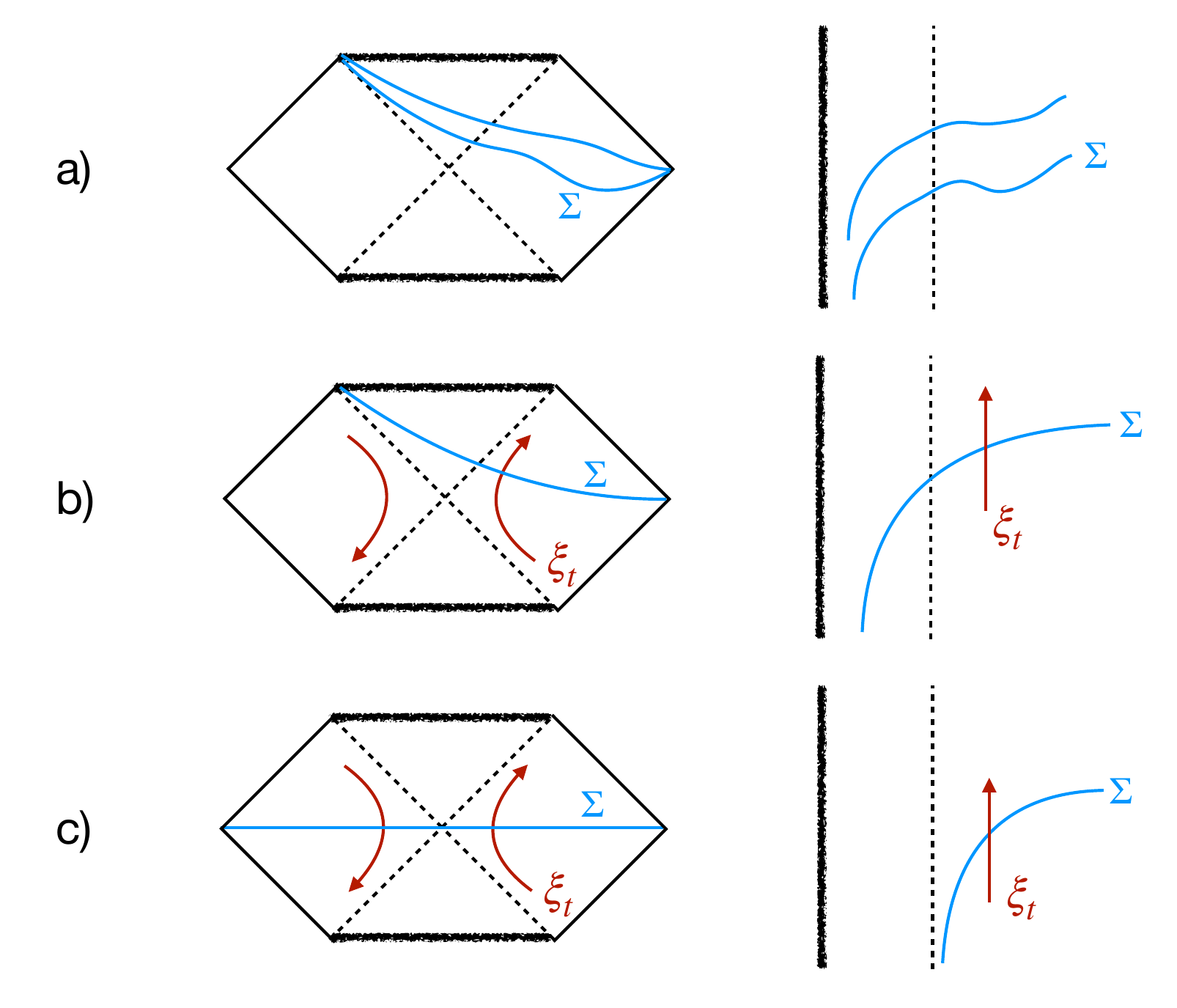} 
 		\caption{The spacetime of an eternal black hole, represented via Penrose or Eddington-Finkelstein diagrams, together with different spatial slicings.  The top row shows a general slicing of the interior and right exterior.  The second row shows one member of a family of stationary slices\cite{BHQIUE,NVU}; other members are found through translation by the Killing vector $\xi_t$.  The third row is a member of a family of extremal slices, also related by translation by the Killing vector.}
 		\label{slices}
 	\end{center}
 \end{figure}

To provide a dressing construction based on the canonical approach described above, one needs a slicing of the spacetime.  There are of course infinitely many such slicings one could consider; for examples see Fig.~\ref{slices}.  For the generic slices shown in Fig.~\ref{slices}a, or the stationary slices shown in Fig.~\ref{slices}b, time evolution generically propagates operators across the horizon; in addition, the generic slicing leads to an explicitly time-dependent background and corresponding additional challenges with describing the unitary evolution.\footnote{See \cite{GiPe3} for a recent overview of the problem, and further references.}

\begin{figure}[h]
 	\begin{center}
 		\includegraphics[width=0.40\textwidth]{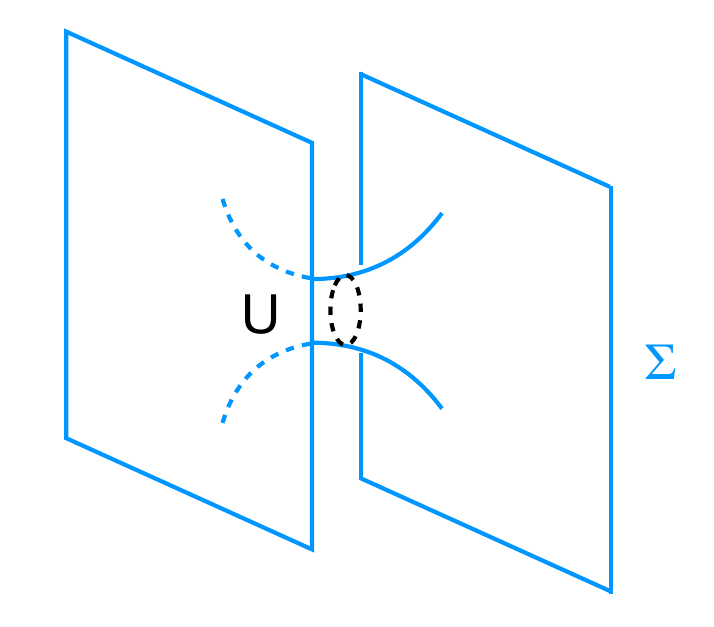} 
 		\caption{The spatial geometry of an extremal slice of Fig.~\ref{slices}c.}
 		\label{slice}
 	\end{center}
 \end{figure} 

The extremal slices of Fig.~\ref{slices}c leads to additional simplification, exploited in \cite{Wittcross,CPW}.  Specifically, we can think of the region $U$ of interest as that   ``inside the horizon."  Of course, as also illustrated in Fig.~\ref{slice}, this really includes an additional asymptotic region.  

The Green function problem of \eqref{GFeqns} can then be explored in this context.  The Schwarzschild metric 
\beq
ds^2=-(1-R/r)dt^2 + dr^2/(1-R/r) + r^2 d\Omega^2\ ,
\eeq
with $R=2GM$,
has a coordinate singularity in the spatial metric at $r=R$.  This may be remedied by defining a new coordinate 
\beq
y= r\sqrt{1-\frac{R}{r} } +\frac{R}{2}\ln\left(\frac{1+ \sqrt{1-\frac{R}{r} }}{1- \sqrt{1-\frac{R}{r} }}\right)\ ,
\eeq
yielding the metric 
\beq\label{newsch}
ds^2=-[1-R/r(y)] dt^2 + dy^2 + r^2(y) d\Omega^2\ .
\eeq
Now the horizon is at $y=0$, and at infinity $y\simeq r$.  The ``interior region" $U$ corresponds to $y<0$, and translation by  $\xi^\mu$ leaves it invariant.  In these coordinates, $\xi^t=1$ and the unit normal is given by $n^t = \pm1/\sqrt{1-R/r(y})$, with sign given by that of $y$, as can easily be seen by comparing with the local flat space limit.

The corresponding time translation generator of an underlying field theory with energy momentum tensor $T_{\mu\nu}$ is thus 
\beq
P_t= -\int d^3y \sqrt{q} \xi^\mu n^\nu T_{\mu\nu}\ ,
\eeq
and therefore acts as an automorphism of the algebra of observables  ${\cal A}_L$ localized to $U$.   That suggests the possibility that the part of the dressing coupling to $P_t$ can yield a bona fide crossed product.

What remains is to separate the dressing into appropriate terms, in an analogy to the standard dressing construction.  Full study of this problem is deferred for future work.  However, for the purposes of making contact with the crossed product, a useful separation is to take
\beq\label{Schsep}
H^0_{ij}(x',x) = H^0_{RS,ij}(x') + \Delta H^0_{ij}(x',x)\ ,
\eeq
where $\kappa H^0_{RS,ij}(x')$ is, for $y>0$, the metric deformation corresponding to a unit-mass deformation of Schwarzschild, $\partial_M q_{ij}$ from \eqref{newsch}, and vanishes for $y<0$.  
This can be thought of as providing a (right) standard dressing associated with the horizon; specifically, when acted on with the derivative operators of \eqref{GFeqns}, it will have a source localized at the horizon, $y=0$.  The remaining term $\Delta H$ can be thought of as providing a ``dressing to the horizon," together with an analog of the angular momentum piece in \eqref{stdress}. For the full dressing one also requires $G^{ijk}$; $H^k_{ij}$ and $G^{ij0}$ can be taken to vanish due to vanishing of the background extrinsic curvature $K_{ij}$ of the extremal slices.

With this decomposition of the Green function, the dressing \eqref{genv} becomes
\bea
V^0(x) &=& V^0_{RS} +\Delta V^0(x)= \frac{\kappa}{2} \int d^3x' H^0_{S,ij}(x') p^{ij}(x') +  \frac{\kappa}{2}\int d^3x' \Delta H^\mu_{ij}(x',x) p^{ij}(x')\cr
V^k(x) &=& \frac{\kappa}{2} \int d^{3}x' G^{ijk}(x',x) h_{ij}(x')\ .
\eea
The exponents in \eqref{genlin} then become
\beq\label{dressdecomp}
\int d^3 x \sqrt q n^\nu V^\mu(x) T_{\mu\nu} = V_{RS}^0 \int  d^3 x \sqrt q \xi^\mu n^\nu T_{\mu\nu} + \int d^3 x \sqrt q n^\mu \left[ \Delta V^0 T_{0\mu} + V^k(x)T_{k\mu}\right]\ .
\eeq

This clearly illustrates a reduction to the crossed product as a {\it truncation} of the more complete gravitational dressing: if the last  terms of eq.~\eqref{dressdecomp} are dropped, what remains is $-V_{RS}^0 P_t$. As noted, $P_t$ is an automorphism --   specifically the modular automorphism associated to the Hartle-Hawking vacuum -- of the subalgebra ${\cal A}_L$ of operators inside the horizon, and since the spacetime is asymptotically flat, the standard dressing $V^0_S$ is again conjugate to the corresponding (right) ADM momentum,
\beq
[P_{R0}^{ADM},V^0_{RS}]=i\ .
\eeq
Then, the truncated analogues of the dressed operators \eqref{Sdressop},
\beq\label{truncop}
\hat O_{trunc}= e^{-iV_{RS}^0P_t} O e^{iV_{RS}^0P_t}
\eeq
 match the crossed product form, \eqref{xprodalg}, involving the modular automorphism.

Various observations can now be made.  First is the question of what algebra or states enter the crossed product construction.  In the way it has been described here, the states and algebra correspond to excitations to the left of the horizon of the eternal black hole, dressed to the right.  This and related algebras were discussed in \cite{CPW}; the related algebras, which also can be constructed via the standard dressing, include $\cala_R$, ``dual" to {\it right} operators dressed to the right, or in an analogous construction with a left dressing, $\cala_L$ ``dual" to left operators dressed to the left. Switching between right and left truncated dressing can be accomplished by subtracting $\partial_M q_{ij}$ from $\kappa H^0_{ij}$, as in \eqref{Schsep}.  In an alternate construction, one could begin with right operators fully dressed to the right, and then make the same subtraction, so that the dressing has nonzero commutator with the left rather than the right ADM hamiltonian; while perhaps more contrived, this also appears to yield a crossed product without truncation.\footnote{I thank Xi Dong for a discussion on this.}  Note also that $V^0_{RS}$ provides a realization in gravitational varaibles of the ``time-shift" operator described in \cite{CPW}; indeed, compare \eqref{dressedO} and \eqref{pertdress}.

Second, it should be clear from this construction that the crossed product truncation is not gauge invariant under the full set of (linearized) diffeomorphisms.  For gauge invariance to hold, as described by commuting with the full set of constraints, the full dressing in \eqref{dressdecomp} must be included\cite{GiPe2}.  So, the commutators of the constraints \eqref{constraints} with the operators  \eqref{truncop} with the truncated dressing will be non-vanishing, associated, {\it e.g.}, with the fact that other of the diffeomorphisms have the effect of moving the neighborhood -- here the black hole interior.  So, in this context, the crossed product construction embeds as a small piece of a bigger algebraic structure involving the more general gravitational dressing.  

Third, one may want to consider a more direct description of the states {\it interior} to the black hole, {\it e.g.} as for a black hole formed from collapse.   The obvious way to proceed is to consider other slicings, such as the stationary slicings of Fig.~\ref{slices}b, or the more general ones of Fig.~\ref{slices}a.  The general construction of the leading dressing described in \cite{GiPe2} can then be used.  However, in these cases there is {\it not} a clear reduction to the bona fide crossed product, since {\it e.g} time translations now move states from exterior to interior of the black hole.  One seemingly must contend directly with the more complete algebraic structure.

The dressing descriptions  of this note clearly provides a more general framework than previously considered for standard dressing constructions.  
Various possible extensions of this discussion apparently exist.  For example, Ref.~\cite{JSS} described obtaining the crossed product and a type II algebra for a region of flat space.  It seems possible to likewise derive this by a similar decomposition of the Green functions associated to the dressing, followed by a truncation of the dressing.  As a result, it appears clear that the construction of \cite{JSS} is likewise not fully gauge invariant, in that the corresponding operators will not commute with the full set of constraints, which, {\it e.g.}, move the region in question.

There is also a direct connection between the soft charge story\cite{Astrosoft} and gravitational dressings\cite{DoGi3,DoGi4,SGsplit}, which one can see should extend to the present context.  Specifically, the choice of particular boundary conditions at infinity for the Green functions $G$ and $H$ includes specification of definite soft charges of the corresponding gravitational field.  We therefore expect to be able to choose different boundary conditions, and thus different $G$ and $H$, for different asymptotics correspond to different soft charges.  These differences could, for example, be incorporated in the choice of different standard dressings.

It also seems clear that such a standard dressing construction can be used in other gauge theories, for example QED.  There one has a simpler Green function problem associated with the dressing, {\it e.g.} as explored in \cite{SGsplit,SGgenasy}, and a simpler problem of decomposing the Green function or dressing into a standard one and the remaining piece.

Of course, the leading gravitational dressing of this paper is expected to be just an approximate ``weak-field" piece of a more complete mathematical structure.  One outstanding problem is to find higher order perturbative corrections to the dressing, and to understand their behavior and further implications.  Going beyond this is the question of construction of the full gauge-invariant gravitationally dressed operators, {\it e.g.} at  nonperturbative level.  There are arguments that their behavior will have important implications in quantum gravity\cite{obs-over}.  For example, one expects that conjugating such an operator by the {\it asymptotic} (or ``boundary") operator
\beq
e^{ia^i P_i^{ADM}}
\eeq
would move an operator associated with a region $U$ to its translate by $a^i$\cite{DoGi3} -- and can translate the operator all the way to the asymptotic region, for large $a^i$.  This provides a possible ``explanation"\cite{DoGi3,SGholo} of the holographic behavior of gravity, simplifying a previous argument due to Marolf\cite{maroholo} involving the boundary hamiltonian $P_0^{ADM}$ -- but one that only applies once one has solved the constraints or their nonperturbative analog\cite{SGholo}.  

Beyond this, the expected behavior of such fully dressed operators would seem to provide a challenge to the utility of an algebraic approach.  To see this, consider a LQFT operator $O_f$ that creates a particle in a localized wavepacket with profile $f$.  Then, the fully dressed operator $\hat O_f$ is expected to also create the gravitational field of that particle, which is asymptotically well-described by the leading dressing discussed in this paper.  However, the operator $\hat O_f^N$ then creates $N$ such particles, and their gravitational fields.  For large $N$, this operator is expected to create something like a large quantum black hole -- which consumes all of spacetime in the limit $N\rightarrow\infty$.  

Such ``algebraic spacetime disruption\cite{obs-over}" suggests that other non-algebraic structure, {\it e.g.} as tentatively investigated in \cite{QFG,QGQF}, might be important in describing localization of information (at least approximately) in quantum gravity.  Specifically, since by this argument an algebra or subalgebra of dressed observables itself seems to describe quantities extending well outside any given region, it may be that some distinct mathematical structure that is not such an algebra plays an important role in characterizing, at least in an approximation, subsystems that we may wish to think of as localized, and provides a foundation for definition of other associated notions such as entanglements and entropies.  
The dressed operators of this and related papers, and other geometric notions, would need to arise in approximation to that more basic structure, and therefore the behavior of these dressed operators is expected to provide important clues about that structure.
However, further discussion of these questions is beyond the scope of the present paper.

\vskip.3in
\noindent{\bf Acknowledgements} 

This material is based upon work supported in part by the U.S. Department of Energy, Office of Science, under Award Number {DE-SC}0011702, and by Heising-Simons Foundation grants \#2021-2819 and \#2024-5307.

\mciteSetMidEndSepPunct{}{\ifmciteBstWouldAddEndPunct.\else\fi}{\relax}
\bibliographystyle{utphys}
\bibliography{ii-beyond}{}

\end{document}